# Design and frequency analysis of continuous finite-time-convergent differentiator

Xinhua Wang and Hai Lin

Department of Electrical & Computer Engineering

National University of Singapore, 4 Engineering Drive 3, Singapore 117576

wangxinhua04@gmail.com

**Abstract:** In this paper, a continuous finite-time-convergent differentiator is presented based on a strong Lyapunov function. The continuous differentiator can reduce chattering phenomenon sufficiently than normal sliding mode differentiator, and the outputs of signal tracking and derivative estimation are all smooth. Frequency analysis is applied to compare the continuous differentiator with sliding mode differentiator. The beauties of the continuous finite-time-convergent differentiator include its simplicity, restraining noises sufficiently, and avoiding the chattering phenomenon.

## 1. Introduction

Differentiation of signals is a well-known problem [1-8], which has attracted much attention in recent years. Obtaining the velocities of tracked targets is crucial for several kinds of systems with correct and timely performances, such as the missile-interception systems [9] and underwater vehicle systems [10], in which disturbances must be restrained. The simpleness and robustness of differentiators should be taken into consideration.

The popular high-gain differentiators [4, 5, 6] provide for an exact derivative when their gains tend to infinity. Unfortunately, their sensitivity to small high-frequency noise also infinitely grows. With any finite gain values such a differentiator has also a finite bandwidth. Thus, being not exact, it is, at the same time, insensitive with respect to high-frequency noise. Such insensitivity may be considered both as advantage or disadvantage depending on the circumstances. Moreover, high gain results in peaking phenomenon.

In [7, 8], a differentiator via second-order (or high-order) sliding modes algorithm was proposed. The information one needs to know on the signal is an upper bound for Lipschitz constant of the derivative of the signal. Although second-order sliding mode is introduced and there exists no chattering phenomenon in signal tracking, derivative estimation still exist chattering phenomenon.

In [11], we presented a finite-time-convergent differentiator based on finite-time stability [12-14] and singular perturbation technique [15-20]. The merits of this differentiator exist in: rapidly finite-time convergence compared with other typical differentiators; no chattering phenomenon. However, peaking phenomenon exist in the output of derivative estimation. Moreover, the ability of restraining noise was not considered.

In [21], we designed a hybrid differentiator with high speed convergence, and it succeeds in applications to velocity estimations for low-speed regions only based on position measurements [22] and to quad-rotor aircraft [23], in which only the convergence of signal tracking was described, but the convergence of derivative estimation was not given, and the regulation of parameters was complex.

In this paper, a continuous finite-time-convergent differentiator is presented based on a strong Lyapunov function. For this differentiator we study its stability and finite time convergence characteristics by means of Lyapunov functions. The continuous differentiator can reduce chattering phenomenon sufficiently than sliding mode differentiator. The advantage of the use of Lyapunov functions is to easily obtain the parameters of differentiator.

Frequency analysis is applied to compare the continuous finite-time-convergent differentiator with sliding mode differentiator. For nonlinear differentiators, an extended version of the frequency response method, describing function method [19, 24], can be used to approximately analyze and predict nonlinear behaviors of nonlinear differentiators. Even though it is only an approximation method, the desirable properties it inherits from the frequency response method, and the shortage of other, systematic tools for nonlinear differentiator analysis, make it an indispensable component of the bag of tools of practicing control engineers. By describing function method, we can conclude that the continuous finite-time-convergent differentiator has the better ability of restraining high-frequency noises than sliding mode differentiator.

This paper is organized in the following format. In section 2, preliminaries are introduced. In section 3, problem statement of sliding mode differentiator is given. In section 4, continuous finite-time-convergent differentiator is presented, and its robustness analysis is given in section 5. In section 6, frequency analysis of continuous differentiator is given. In section 7, the simulations are given, and our conclusions are made in section 8.

## 2. Preliminaries

First of all, the concepts related to finite-time control are given (See [13]).

**Definition 1:** Consider a time-invariant system in the form of

$$\dot{x} = f(x), \quad f(0) = 0, \quad x \in R^n \quad (1)$$

where $f : \hat{U}_0 \to R^n$ is continuous on an open neighborhood $\hat{U}_0$ of the origin. The equilibrium $x = 0$ of the system is (locally) finite-time stable if (i) it is asymptotically stable, in $\hat{U}$, an open neighborhood of the origin, with $\hat{U} \subseteq \hat{U}_0$; (ii) it is finite-time convergent in $\hat{U}$, that is, for any initial condition $x_0 \in \hat{U} \setminus \{0\}$, there is a settling time $T > 0$ such that every solution $x(t, x_0)$ of system (1) is defined with $x(t, x_0) \in \hat{U} \setminus \{0\}$ for $t \in [0, T]$ and satisfies

$$\lim_{t \to T} x(t, x_0) = 0 \quad (2)$$

and $x(t, x_0) = 0$, if $t \geq T$. Moreover, if $\hat{U} = R^n$, the origin $x = 0$ is globally finite-time stable.

**Definition 2:** A family of dilations $\delta_\rho^r$ is a mapping that assigns to every real $\rho > 0$ a diffeomorphism

$$\delta_\rho^r(x_1, \cdots, x_n) = (\rho^{r_1} x_1, \cdots, \rho^{r_n} x_n) \quad (3)$$

where $x_1, \cdots x_n$ are suitable coordinates on $R^n$ and $r = (r_1, \cdots, r_n)$ with the dilation coefficients $r_1, \cdots, r_n$ positive real numbers.

A vector field $f(x) = (f_1(x), \cdots f_n(x))^T$ is homogeneous of degree $k \in R$ with respect to the family of dilations $\delta_\rho^r$ if

$$f_i(\rho^{r_1} x_1, \cdots, \rho^{r_n} x_n) = \rho^{r_i + k} f_i(x) \quad (4)$$
$$i = 1, \cdots, n, \quad \rho > 0$$

System (1) is called homogeneous if its vector field $f$ is homogeneous.

The following lemma was presented in some references like [12, 25, 26].

**Lemma1:** Suppose that system (1) is homogeneous of degree $k < 0$ with respect to the family of dilations $\delta_\rho^r$, $f(x)$ is continuous and $x = 0$ is its asymptotically stable equilibrium. Then equilibrium of system (1) is globally finite-time stable.

**Lemma 2** [14, Theorem 4.2]. Suppose there exists a continuous function $V: D \to R$ such that the following conditions hold:

(1) $V$ is positive definite;

(2) There exist real number $c > 0$ and $\theta \in (0,1)$ and an open neighborhood $\nu \subset D$ of the origin such that

$$\dot{V}(x) + c(V(x))^\theta \leq 0, \quad x \in \nu \setminus \{0\} \quad (5)$$

Then the origin is a finite-time-stable equilibrium of (1), and the settling-time function $T$ is

$$T(x) \leq \frac{1}{c(1-\theta)}(V(x))^{1-\theta} \quad (6)$$

and $T$ is continuous. If in addition $D = R^n$, $V$ is proper, and $\dot{V}$ takes negative values on $R^n \setminus \{0\}$, then the origin is a globally finite-time-stable equilibrium of (1).

**Assumption 1.** For (1), there exist $\rho \in (0,1]$ and a nonnegative constant $\bar{a}$ such that

$$\|f(z_1) - f(\bar{z}_1)\| \leq \bar{a} \|z_i - \bar{z}_i\|^\rho \quad (7)$$

where $z, \bar{z} \in \Re^n$.

**Remark 1.** There are a number of nonlinear functions actually satisfying Assumption 1. For example, one such function is $x^\rho$ since $|x^\rho - \bar{x}^\rho| \leq 2^{1-\rho}|x-\bar{x}|^\rho$, $\rho \in (0,1]$. Moreover, there are smooth functions also satisfying this property. In fact, it is easy to verify that $|\sin x - \sin \bar{x}| \leq 2|x-\bar{x}|^\rho$ for any $\rho \in (0,1]$.

## 3. Problem statement of sliding mode differentiator
### 3.1 Stability analysis of sliding mode differentiator

In [7, 8], a finite-time convergent differentiator (Levant differentiator) based on second-order sliding is presented as follow:

$$\begin{aligned}\dot{x}_1 &= x_2 - k_1 |x_1 - v(t)|^{\frac{1}{2}} \text{sgn}(x_1 - v(t)) \\ \dot{x}_2 &= -k_2 \text{sgn}(x_1 - v(t))\end{aligned} \quad (8)$$

where the second-order derivative of signal $v(t)$ is bounded, and $|\ddot{v}(t)| \leq L_2$, $L_2$ is a positive constant, $k_1, k_2 > 0$, $k_2 > L_2$. For differentiator (8), there exists a time $t_s > 0$ such that

$$x_1 = v(t), x_2 = \dot{v}(t) \quad (9)$$

for $t \geq t_s$. In fact, we let

$$e_1 = x_1 - v(t), e_2 = x_2 - \dot{v}(t) \quad (10)$$

The tracking error system is

$$\begin{aligned}\dot{e}_1 &= e_2 - k_1 |e_1|^{0.5} \text{sgn}(e_1) \\ \dot{e}_2 &= -k_2 \text{sgn}(e_1) - \ddot{v}(t)\end{aligned} \quad (11)$$

We can get the following differential inclusion:

$$\begin{aligned}\dot{e}_1 &= e_2 - k_1 |e_1|^{0.5} \text{sgn}(e_1) \\ \dot{e}_2 &\in -[k_2 - L_2, k_2 + L_2]\text{sgn}(e_1)\end{aligned} \quad (12)$$

or

$$\begin{aligned}\dot{e}_1 &= e_2 - k_1 |e_1|^{0.5} \text{sgn}(e_1), \\ \dot{e}_2 &= -\bar{k}_2 \text{sgn}(e_1)\end{aligned} \bar{k}_2 \in [k_2 - L_2, k_2 + L_2] \quad (13)$$

Let the Lyapunov function be

$$V = \bar{k}_2 |e_1| + \frac{1}{2} e_2^2 \quad (14)$$

Therefore,

$$\begin{aligned}\dot{V} &= \bar{k}_2 \text{sgn}(e_1)\left(e_2 - k_1 |e_1|^{0.5} \text{sgn}(e_1)\right) + e_2\left(-\bar{k}_2 \text{sgn}(e_1)\right) \\ &= -k_1 \bar{k}_2 |e_1|^{0.5} < 0\end{aligned} \quad (15)$$

From the Lasell's invariant set, the error system is uniformly asymptotically stable. From Definition 2, let

$$\rho^{r_2} e_2 - k_1 |\rho^{r_1} e_1|^{0.5} \text{sgn}(e_1) = \rho^{r_1+k}\left(e_2 - k_1 |e_1|^{0.5} \text{sgn}(e_1)\right)$$

and

$$\begin{aligned}&-[k_2 - L_2, k_2 - L_2]\text{sgn}(\rho^{r_1} e_1) \\ &= \rho^{r_2+k}\left(-[k_2 - L_2, k_2 - L_2]\text{sgn}(e_1)\right)\end{aligned} \quad (16)$$

Therefore, we have

$$r_2 = 0.5 r_1 = r_1 + k, 0 = r_2 + k \quad (17)$$

and we can get

$$k < 0 \quad (18)$$

Therefore, from Lemma 1, differentiator (8) is finite-time stable.

### 3.2 Robustness analysis of sliding mode differentiator

Moreover, let $v(t)$ is the input signal with noise, $v_0(t)$ is the desired signal, and it is satisfied with $|v(t) - v_0(t)| \leq \sigma$. Therefore, for some positive constants $\mu_1$ and $\mu_2$ the following inequalities are established [8]:

$$|e_1| = |x_1 - v_0(t)| \leq \mu_1 \sigma, |e_2| = |x_2 - \dot{v}_0(t)| \leq \mu_2 \sigma^{\frac{1}{2}} \quad (19)$$

### 3.3 Chattering phenomenon

Because switch function exists in the second differential equation of differentiator (8), although the output $x_1$ is smooth, the output $x_2$ is continuous but non-smooth, it is called as chattering phenomenon. In order to explain the problem, we give an example in the following.

**Example 1:** For system

$$\begin{aligned}\dot{x}_1 &= x_2 - k_1 |x_1|^{\frac{1}{2}} \text{sgn}(x_1) \\ \dot{x}_2 &= -k_2 \text{sgn}(x_1)\end{aligned} \quad (20)$$

let $k_1 = 6, k_2 = 9$, then we have the solutions $x_1$ and $x_2$ in Fig. 1.

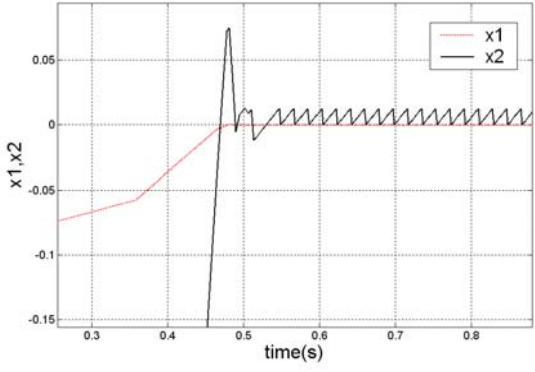

Fig. 1 $x_1$ and $x_2$ of system (20)

From Fig. 1, we find that, for $x_2$, chattering phenomenon happens near the equilibrium. Moreover, if noises exist in signal, this chattering phenomenon will magnify noises near the equilibrium. In some velocity feedback systems, this chattering in $x_2$ can make motors trembling. Therefore, chattering phenomenon must be removed sufficiently in the output $x_2$ of a differentiator.

**3.4 Frequency analysis of sliding mode differentiator**

We can also analyse sliding differentiator (8) from frequency characteristics.

Let $e_1 = x_1 - v(t) = A\sin\omega t$, we have

$$\frac{2}{\pi}\int_0^\pi |A\sin\omega\tau|^{0.5}\operatorname{sgn}(A\sin\omega\tau)\sin\omega\tau d\omega t \qquad (21)$$
$$= A^{0.5}\frac{2}{\pi}\int_0^\pi |\sin\omega\tau|^{1.5} d\omega t$$

Then we can get

$$\frac{2}{\pi}\int_0^\pi \sin\omega\tau d\omega t = \frac{2}{\pi}(-\cos\omega\tau)\big|_0^\pi = \frac{4}{\pi} \qquad (22)$$

and

$$\frac{2}{\pi}\int_0^\pi (\sin\omega\tau)^2 d\omega t = \frac{2}{\pi}\int_0^\pi \frac{1-\cos 2\omega\tau}{2} d\omega t = 1 \qquad (23)$$

In (21), let $\Omega = \frac{2}{\pi}\int_0^\pi |\sin\omega\tau|^{1.5} d\omega t$, we have

$$1 < \Omega < \frac{4}{\pi} \qquad (24)$$

Therefore, the describing function of nonlinear function $|\ |^{0.5}\operatorname{sgn}(\ )$ is

$$N(A) = \frac{\Omega}{A^{0.5}} \qquad (25)$$

and the linearization system of differentiator (8) is

$$\dot{x}_1 = x_2 - \lambda_2 \frac{\Omega}{A^{0.5}}(x_1 - v(t)) \qquad (26)$$
$$\dot{x}_2 = -\lambda_1 \frac{4}{\pi A}(x_1 - v(t))$$

The nature frequency of system (26) is

$$\omega_n = \frac{2\sqrt{k_2}}{\sqrt{\pi}A^{0.5}} \qquad (27)$$

and we have

$$2\varsigma\omega_n = \frac{k_1\Omega}{A^{0.5}} \qquad (28)$$

Therefore, the damping coefficient is

$$\varsigma = \frac{k_1\Omega\sqrt{\pi}}{4\sqrt{k_2}} \qquad (29)$$

With the error amplitude decreasing, the nature frequency $\omega_n$ increases. Moreover, chattering phenomenon happen rear the equilibrium for the discontinuous differentiator. If noises exist in signal, this chattering phenomenon will magnify noises.

In order to remove chattering phenomenon and to restrain sufficiently high-frequency noises, we should design a continuous differentiator.

**4. Continuous finite-time-convergent differentiator**

**4.1 Continuous finite-time-convergent system**

In order to design continuous differentiator, firstly, we give a finite-time stability Theorem as follow.

**Theorem 1:** For continuous system

$$\dot{z}_1 = z_2 - k_1|z_1|^{\frac{\alpha+1}{2}}\operatorname{sgn}(z_1) \qquad (30)$$
$$\dot{z}_2 = -k_2|z_1|^\alpha \operatorname{sgn}(z_1)$$

there exist constants $k_1, k_2 > 0$, $t_s > 0$ and $\alpha \in (0,1)$ such that $z_1$ and $z_2$ are smooth, and

$$z_1 = 0, z_2 = 0 \qquad (31)$$

for $t \geq t_s$, i.e., system (30) is finite-time stable with respect to the origin.

**Proof:** We select a Lyapunov function as

$$V = \frac{2k_2}{\alpha+1}|z_1|^{\alpha+1} + \frac{1}{2}z_2^2 + \frac{1}{2}\left(k_1|z_1|^{\frac{\alpha+1}{2}}\operatorname{sgn}(z_1) - z_2\right)^2 \qquad (32)$$

and we can get

$$V = \left(\frac{2k_2}{\alpha+1} + \frac{1}{2}k_1^2\right)|z_1|^{\alpha+1} + z_2^2 - k_1 z_2 |z_1|^{\frac{\alpha+1}{2}} \operatorname{sgn}(z_1)$$

$$= \begin{bmatrix} |z_1|^{\frac{\alpha+1}{2}} \operatorname{sgn}(z_1) & z_2 \end{bmatrix} \frac{1}{2} \begin{bmatrix} \frac{4k_2}{\alpha+1} + k_1^2 & -k_1 \\ -k_1 & 2 \end{bmatrix} \quad (33)$$

$$\times \begin{bmatrix} |z_1|^{\frac{\alpha+1}{2}} \operatorname{sgn}(z_1) \\ z_2 \end{bmatrix}$$

Let

$$\varsigma = \begin{bmatrix} |z_1|^{\frac{\alpha+1}{2}} \operatorname{sgn}(z_1) & z_2 \end{bmatrix}^T \quad (34)$$

$$P = \frac{1}{2} \begin{bmatrix} \frac{4k_2}{\alpha+1} + k_1^2 & -k_1 \\ -k_1 & 2 \end{bmatrix} \quad (35)$$

we have

$$V = \varsigma^T P \varsigma \quad (36)$$

From (35), we know that matrix $P$ is symmetrical and positive-definite, and

$$\lambda_{\min}\{P\} \|\varsigma\|_2^2 \leq V \leq \lambda_{\max}\{P\} \|\varsigma\|_2^2 \quad (37)$$

is satisfied, where

$$\|\varsigma\|_2^2 = |z_1|^{\alpha+1} + z_2^2 \quad (38)$$

The time derivative of Lyapunov function (32) along the solutions of system (30) is

$$\dot{V} = -2k_1 k_2 |z_1|^{\frac{3\alpha+1}{2}} + \frac{k_1^2(\alpha+1)}{2} z_2 |z_1|^{\alpha} \operatorname{sgn}(z_1)$$

$$+ k_1 \left(k_2 - \frac{k_1^2(\alpha+1)}{2}\right) |z_1|^{\frac{3\alpha+1}{2}} - \frac{k_1(\alpha+1)}{2} z_2^2 |z_1|^{\frac{\alpha-1}{2}} \quad (39)$$

$$+ \frac{k_1^2(\alpha+1)}{2} z_2 |z_1|^{\alpha} \operatorname{sgn}(z_1)$$

Therefore, we have

$$\dot{V} = -|z_1|^{\frac{\alpha-1}{2}} \begin{bmatrix} |z_1|^{\frac{\alpha+1}{2}} \operatorname{sgn}(z_1) & z_2 \end{bmatrix} \frac{k_1}{2}$$

$$\times \begin{bmatrix} 2k_2 + k_1^2(\alpha+1) & -k_1(\alpha+1) \\ -k_1(\alpha+1) & (\alpha+1) \end{bmatrix} \begin{bmatrix} |z_1|^{\frac{\alpha+1}{2}} \operatorname{sgn}(z_1) \\ z_2 \end{bmatrix} \quad (40)$$

Let

$$Q = \frac{k_1}{2} \begin{bmatrix} 2k_2 + k_1^2(\alpha+1) & -k_1(\alpha+1) \\ -k_1(\alpha+1) & (\alpha+1) \end{bmatrix} \quad (41)$$

we can get

$$\dot{V} = -|z_1|^{\frac{\alpha-1}{2}} \varsigma^T Q \varsigma \leq -|z_1|^{\frac{\alpha-1}{2}} \lambda_{\min}\{Q\} \|\varsigma\|_2^2 \quad (42)$$

From (37), we can get

$$|z_1|^{\alpha+1} \leq \|\varsigma\|_2^2 \leq \frac{V}{\lambda_{\min}\{P\}} \quad (43)$$

Therefore, we have

$$|z_1|^{\frac{\alpha-1}{2}} \geq \left(\frac{V}{\lambda_{\min}\{P\}}\right)^{\frac{\alpha-1}{2(\alpha+1)}} \quad (44)$$

Then

$$\dot{V} \leq -\frac{(\lambda_{\min}\{P\})^{\frac{1-\alpha}{2(\alpha+1)}} \lambda_{\min}\{Q\}}{\lambda_{\min}\{P\}} V^{\frac{3\alpha+1}{2(\alpha+1)}} \quad (45)$$

Because

$$\frac{3\alpha+1}{2(\alpha+1)} = \frac{3\alpha+1}{3\alpha+1+(1-\alpha)} \quad (46)$$

we have

$$0 < \frac{3\alpha+1}{2(\alpha+1)} < 1 \quad (47)$$

From (45), (47) and Lemma 2, system (30) is uniformly finite-time stable. ∎

In order to explain the smoothness of system solutions, we give an example in the following.

**Example 2:** For system (30), let $k_1 = 6, k_2 = 9, \alpha = 0.2$, then we have the solutions $x_1$ and $x_2$ in Fig. 2.

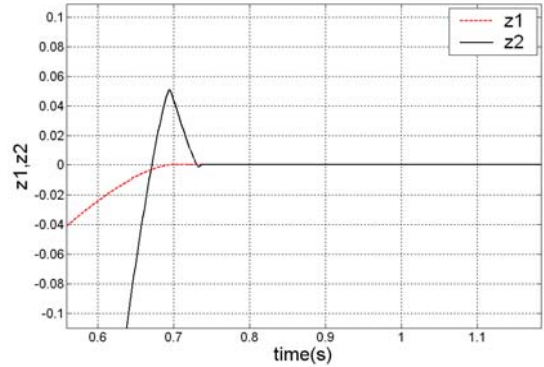

Fig. 2 z1 and z2 of system (30)

From Fig. 2, we find that the solutions $z_1$ and $z_2$ of system (30) are all smooth, no chattering phenomenon happen.

**4.2 Design of continuous finite-time-convergent differentiator**

In the following, we will present a continuous finite-time-convergent differentiator.

**Theorem 2:** For continuous differentiator

$$\dot{x}_1 = x_2 - k_1 |x_1 - v(t)|^{\frac{\alpha+1}{2}} \operatorname{sgn}(x_1 - v(t))$$
$$\dot{x}_2 = -k_2 |x_1 - v(t)|^{\alpha} \operatorname{sgn}(x_1 - v(t)) \quad (48)$$

and second-order derivable signal $v(t)$, there exist constants $k_1, k_2 > 0$, $t_s > 0$ and $0 < \alpha < 1$ such that the outputs $x_1$ and $x_2$ of differentiator (48) are smooth, and

$$\|\varsigma\|_2 \leq \left(\frac{lL_2}{\lambda_{\min}\{Q\}}\right)^{\frac{\alpha+1}{2\alpha}} \quad (49)$$

is satisfied. Moreover, when $\alpha = 0$, output $x_1$ is smooth and $x_2$ is continuous but non-smooth, and $\varsigma = 0$ for $t \geq t_s$. Where $|\ddot{v}(t)| \leq L_2$, and

$$\varsigma = \left[|e_1|^{\frac{\alpha+1}{2}} \operatorname{sgn}(e_1) \quad e_2\right]^T, \quad e_1 = x_1 - v(t),$$

$$e_2 = x_2 - \dot{v}(t), \quad (50)$$

$$l = \|[k_1 \quad -2]\|_2,$$

$$Q = \frac{k_1}{2}\begin{bmatrix} 2k_2 + k_1^2(\alpha+1) & -k_1(\alpha+1) \\ -k_1(\alpha+1) & \alpha+1 \end{bmatrix}, \quad (51)$$

$$k_1 > 0, \quad k_2 > \frac{2(k_1^2+4)L_2^2}{k_1^2(\alpha+1)} \quad (52)$$

$\frac{lL_2}{\lambda_{\min}\{Q\}} < 1$ can be obtained in differentiator (48) and $0 < \alpha < 1$ is selected sufficiently small, therefore, $\|\varsigma\|_2$ becomes sufficiently small in a finite time. This will be given in the following proof.

Comparing with discontinuous differentiator (8), continuous differentiator (48) is continuous and its outputs are all smooth. Moreover, because a strong Lyapunov function is selected to design the differentiator, regulation of parameters is easier.

**Proof:** The error system of differentiator (48) is

$$\dot{e}_1 = e_2 - k_1|e_1|^{\frac{\alpha+1}{2}} \operatorname{sgn}(e_1) \quad (53)$$
$$\dot{e}_2 = -k_2|e_1|^{\alpha} \operatorname{sgn}(e_1) - \ddot{v}(t)$$

Select a Lyapunov function as

$$V = \frac{2k_2}{\alpha+1}|e_1|^{\alpha+1} + \frac{1}{2}e_2^2 + \frac{1}{2}\left(k_1|e_1|^{\frac{\alpha+1}{2}} \operatorname{sgn}(e_1) - e_2\right)^2 \quad (54)$$

The time derivative of Lyapunov function (54) along the solutions of system (53) is

$$\dot{V} = -2k_1k_2|e_1|^{\frac{3\alpha+1}{2}} + \frac{k_1^2(\alpha+1)}{2}e_2|e_1|^{\alpha}\operatorname{sgn}(e_1)$$
$$+ k_1\left(k_2 - \frac{k_1^2(\alpha+1)}{2}\right)|e_1|^{\frac{3\alpha+1}{2}} - \frac{k_1(\alpha+1)}{2}e_2^2|e_1|^{\frac{\alpha-1}{2}} \quad (55)$$
$$+ \frac{k_1^2(\alpha+1)}{2}e_2|e_1|^{\alpha}\operatorname{sgn}(e_1)$$
$$+ \left(k_1|e_1|^{\frac{\alpha+1}{2}}\operatorname{sgn}(e_1) - e_2\right)\ddot{v}(t) - e_2\ddot{v}(t)$$

Therefore, we can get

$$\dot{V} = -|e_1|^{\frac{\alpha-1}{2}}\left[|e_1|^{\frac{\alpha+1}{2}}\operatorname{sgn}(e_1) \quad e_2\right]\frac{k_1}{2}$$
$$\times \begin{bmatrix} 2k_2 + k_1^2(\alpha+1) & -k_1(\alpha+1) \\ -k_1(\alpha+1) & \alpha+1 \end{bmatrix}\begin{bmatrix}|e_1|^{\frac{\alpha+1}{2}}\operatorname{sgn}(e_1) \\ e_2\end{bmatrix} \quad (56)$$
$$+ [k_1 \quad -2]\begin{bmatrix}|e_1|^{\frac{\alpha+1}{2}}\operatorname{sgn}(z_1) \\ e_2\end{bmatrix}\ddot{v}(t)$$

Let

$$\varsigma = \left[|e_1|^{\frac{\alpha+1}{2}}\operatorname{sgn}(e_1) \quad e_2\right]^T \quad (57)$$

$$Q = \frac{k_1}{2}\begin{bmatrix} 2k_2 + k_1^2(\alpha+1) & -k_1(\alpha+1) \\ -k_1(\alpha+1) & \alpha+1 \end{bmatrix} \quad (58)$$

and

$$l = \|[k_1 \quad -2]\|_2 = \sqrt{k_1^2+4} \quad (59)$$

From (57), we have

$$\|\varsigma\|_2^2 = |e_1|^{\alpha+1} + e_2^2 \quad (60)$$

From (60) and $0 < \alpha < 1$, we have

$$|e_1|^{\frac{\alpha-1}{2}} \geq \|\varsigma\|_2^{\frac{\alpha-1}{\alpha+1}} \quad (61)$$

Therefore, we can get

$$\dot{V} \leq -\lambda_{\min}\{Q\}\|\varsigma\|_2^{\frac{\alpha-1}{\alpha+1}}\|\varsigma\|_2^2 + lL_2\|\varsigma\|$$
$$= -\lambda_{\min}\{Q\}\|\varsigma\|_2^{\frac{3\alpha+1}{\alpha+1}} - lL_2\|\varsigma\| \quad (62)$$
$$= -\left(\lambda_{\min}\{Q\}\|\varsigma\|_2^{\frac{2\alpha}{\alpha+1}} - lL_2\right)\|\varsigma\|$$

Moreover, we have

$$\|\varsigma\|_2 \leq \left(\frac{lL_2}{\lambda_{\min}\{Q\}}\right)^{\frac{\alpha+1}{2\alpha}} \quad (63)$$

We can select $k_1$ and $k_2$ such that

$$\lambda_{\min}\{Q\} > lL_2 \quad (64)$$

In fact,

$$|sI - Q| = \begin{vmatrix} s - \frac{k_1}{2}(2k_2 + k_1^2(\alpha+1)) & \frac{k_1}{2}k_1(\alpha+1) \\ \frac{k_1}{2}k_1(\alpha+1) & s - \frac{k_1}{2}(\alpha+1) \end{vmatrix} = 0 \quad (65)$$

i.e.,

$$s^2 - \frac{k_1}{2}(2k_2 + (k_1^2+1)(\alpha+1))s + \frac{k_1^2}{4}2k_2(\alpha+1) = 0 \quad (66)$$

The minimum eigenvalue of (66) is

$$\lambda_{\min}\{Q\} = \frac{k_1}{2}\left( \frac{(2k_2 + (k_1^2+1)(\alpha+1))}{2} - \frac{\sqrt{(2k_2 + (k_1^2+1)(\alpha+1))^2 - 8k_2(\alpha+1)}}{2} \right) \quad (67)$$

Because $\lambda_{\min}\{Q\} > lL_2$ and $k_1 > 0$ are required, we have

$$\frac{k_1}{2}\left( \frac{(2k_2 + (k_1^2+1)(\alpha+1))}{2} - \frac{\sqrt{(2k_2 + (k_1^2+1)(\alpha+1))^2 - 8k_2(\alpha+1)}}{2} \right) > L_2\sqrt{k_1^2+4} \quad (68)$$

Therefore, we get

$$k_2 > \frac{2(k_1^2+4)L_2^2}{k_1^2(\alpha+1)} \quad (69)$$

and

$$k_1 > 0 \quad (70)$$

Therefore, when $0 < \alpha < 1$ is sufficiently small, $\frac{\alpha+1}{2\alpha}$ is sufficiently large, finally, the tracking and estimation errors are sufficiently small in a finite time.

When $\alpha = 0$, Levant differentiator (8) is obtained, and the Lyapunov function can be designed as

$$V = \frac{2k_2}{\alpha+1}|e_1| + \frac{1}{2}e_2^2 + \frac{1}{2}\left(k_1|e_1|^{\frac{1}{2}}\operatorname{sgn}(e_1) - e_2\right)^2 \quad (71)$$

output $x_1$ is smooth and $x_2$ is continuous but non-smooth, and $\varsigma = 0$ for $t \geq t_s$. This concludes the proof. ∎

## 5. Robustness analysis of continuous differentiator

**Theorem 3:** For differentiator (48), if there exist a noise in signal $v(t)$, i.e., $v(t) = v_0(t) + \delta(t)$, where $v_0(t)$ is the desired second-order derivable signal, $\delta(t)$ is a bounded noise and satisfied with $|\delta(t)| \leq \sigma$. Then, the following inequality is established in finite time

$$\|\varsigma\|_2 \leq \frac{\Psi_1(\sigma)}{\lambda_{\min}\{Q\} - l_1 L_2 - \Psi_2(\sigma)} \quad (72)$$

where

$$\varsigma = \left[ |e_1|^{\frac{\alpha+1}{2}}\operatorname{sgn}(e_1) \quad e_2 \right]^T, \quad e_1 = x_1 - v_0(t),$$

$$e_2 = x_2 - \dot{v}_0(t), \quad |\ddot{v}_0(t)| \leq L_2$$

$$l_1 = \|[k_1 \quad -2]\|_2, \quad \Psi_1(\sigma) = k_1\left[k_2 + \frac{k_1(\alpha+1)}{2}l_2\right]2^{\frac{1-\alpha}{2}}\sigma^{\frac{\alpha+1}{2}},$$

$$\Psi_2(\sigma) = k_2[1+l_2]2^{1-\alpha}\sigma^\alpha$$

$$Q = \frac{k_1}{2}\begin{bmatrix} 2k_2 + k_1^2(\alpha+1) & -k_1(\alpha+1) \\ -k_1(\alpha+1) & \alpha+1 \end{bmatrix} \quad (73)$$

**Proof:** Let

$$e_1 = x_1 - v_0(t), e_2 = x_2 - \dot{v}_0(t) \quad (74)$$

The error system is

$$\dot{e}_1 = e_2 - k_1|e_1 - \delta|^{\frac{\alpha+1}{2}}\operatorname{sgn}(e_1 - \delta) \quad (75)$$
$$\dot{e}_2 = -k_2|e_1 - \delta|^\alpha \operatorname{sgn}(e_1 - \delta) - \ddot{v}_0(t)$$

Let

$$\Delta_1 = -|e_1 - \delta|^{\frac{\alpha+1}{2}}\operatorname{sgn}(e_1 - \delta) + |e_1|^{\frac{\alpha+1}{2}}\operatorname{sgn}(e_1) \quad (76)$$
$$\Delta_2 = -|e_1 - \delta|^\alpha \operatorname{sgn}(e_1 - \delta) + |e_1|^\alpha \operatorname{sgn}(e_1)$$

Therefore, we have

$$|\Delta_1| \leq 2^{\frac{1-\alpha}{2}}|\delta|^{\frac{\alpha+1}{2}} \leq 2^{\frac{1-\alpha}{2}}\sigma^{\frac{\alpha+1}{2}} \quad (77)$$
$$|\Delta_2| \leq 2^{1-\alpha}|\delta|^\alpha \leq 2^{1-\alpha}\sigma^\alpha$$

The Lyapunov function is selected as

$$V = \frac{2k_2}{\alpha+1}|e_1|^{\alpha+1} + \frac{1}{2}e_2^2 + \frac{1}{2}\left(k_1|e_1|^{\frac{\alpha+1}{2}}\operatorname{sgn}(e_1) - e_2\right)^2 \quad (78)$$

Let

$$\varsigma = \begin{bmatrix}|e_1|^{\frac{\alpha+1}{2}}\operatorname{sgn}(e_1) & e_2\end{bmatrix}^T, \; P = \frac{1}{2}\begin{bmatrix}\frac{4k_2}{\alpha+1}+k_1^2 & -k_1 \\ -k_1 & 2\end{bmatrix} \quad (79)$$

we can get

$$V = \varsigma^T P \varsigma \quad (80)$$

and we know that matrix $P$ is symmetrical and positive-definite, and

$$\lambda_{\min}\{P\}\|\varsigma\|_2^2 \le V \le \lambda_{\max}\{P\}\|\varsigma\|_2^2 \quad (81)$$

is satisfied, where

$$\|\varsigma\|_2^2 = |e_1|^{\alpha+1} + e_2^2 \quad (82)$$

The time derivative of Lyapunov function (78) along the solutions of error system (75) is

$$\begin{aligned}\dot{V} &= -|e_1|^{\frac{\alpha-1}{2}}\varsigma^T Q\varsigma + [k_1 \;\; -2]\varsigma \ddot{v}(t) \\ &\quad + 2k_1 k_2 \Delta_1 |e_1|^\alpha \operatorname{sgn}(e_1) + k_2 \Delta_2 e_2 \\ &\quad + [k_1 \;\; -1]\varsigma \frac{k_1(\alpha+1)}{2}|e_1|^{\frac{\alpha-1}{2}}k_1\Delta_1 - [k_1 \;\; -1]\varsigma k_2 \Delta_2 \\ &\le -|e_1|^{\frac{\alpha-1}{2}}\lambda_{\min}\{Q\}\|\varsigma\|_2^2 + l_1 L_2 \|\varsigma\|_2 \\ &\quad + |e_1|^{\frac{\alpha-1}{2}}k_1\left[k_2 + \frac{k_1(\alpha+1)}{2}l_2\right]2^{\frac{1-\alpha}{2}}\varepsilon^{\frac{\alpha+1}{2}}\|\varsigma\|_2 \\ &\quad + k_2[1+l_2]2^{1-\alpha}\varepsilon^\alpha\|\varsigma\|_2\end{aligned} \quad (83)$$

where

$$Q = \frac{k_1}{2}\begin{bmatrix}2k_2+k_1^2(\alpha+1) & -k_1(\alpha+1) \\ -k_1(\alpha+1) & \alpha+1\end{bmatrix}, \; l_1 = \|[k_1 \;\; -2]\|_2,$$

$$l_2 = \|[k_1 \;\; -1]\|_2 \quad (84)$$

Let

$$\Psi_1(\sigma) = k_1\left[k_2 + \frac{k_1(\alpha+1)}{2}l_2\right]2^{\frac{1-\alpha}{2}}\sigma^{\frac{\alpha+1}{2}} \quad (85)$$

$$\Psi_2(\sigma) = k_2[1+l_2]2^{1-\alpha}\sigma^\alpha \quad (86)$$

Therefore, we have

$$\begin{aligned}\dot{V} &\le -|e_1|^{\frac{\alpha-1}{2}}\lambda_{\min}\{Q\}\|\varsigma\|_2^2 + l_1 L_2 \|\varsigma\|_2 \\ &\quad + |e_1|^{\frac{\alpha-1}{2}}\Psi_1(\sigma)\|\varsigma\|_2 + \Psi_2(\sigma)\|\varsigma\|_2\end{aligned} \quad (87)$$

Suppose there exist a positive constant $c_1$ such that

$$c_1\|\varsigma\|_2 > \Psi_1(\sigma) \quad (88)$$

Therefore, we have

$$\dot{V} \le -|e_1|^{\frac{\alpha-1}{2}}\left[\lambda_{\min}\{Q\}-c_1\right]\|\varsigma\|_2^2 + \left[l_1 L_2 + \Psi_2(\sigma)\right]\|\varsigma\|_2 \quad (89)$$

From (82) and $0<\alpha<1$, we get

$$|e_1|^{\frac{\alpha-1}{2}} \ge \|\varsigma\|_2^{\frac{\alpha-1}{\alpha+1}} \quad (90)$$

Therefore,

$$\begin{aligned}\dot{V} &\le -\|\varsigma\|_2^{\frac{\alpha-1}{\alpha+1}}\left[\lambda_{\min}\{Q\}-c_1\right]\|\varsigma\|_2^2 + \left[l_1 L_2 + \Psi_2(\sigma)\right]\|\varsigma\|_2 \\ &= -\left\{\left[\lambda_{\min}\{Q\}-c_1\right]\|\varsigma\|_2^{\frac{2\alpha}{\alpha+1}} - \left[l_1 L_2 + \Psi_2(\sigma)\right]\right\}\|\varsigma\|_2\end{aligned} \quad (91)$$

From (81), we have

$$\frac{V}{\lambda_{\max}\{P\}} \le \|\varsigma\|_2^2 \le \frac{V}{\lambda_{\min}\{P\}} \quad (92)$$

From (91) and (92), we can get

$$\dot{V} \le -\left\{\left[\lambda_{\min}\{Q\}-c_1\right]\|\varsigma\|_2^{\frac{2\alpha}{\alpha+1}} - \left[l_1 L_2 + \Psi_2(\sigma)\right]\right\}\frac{V^{0.5}}{\sqrt{\lambda_{\max}\{P\}}} \quad (93)$$

If $\left[\lambda_{\min}\{Q\}-c_1\right]\|\varsigma\|_2^{\frac{2\alpha}{\alpha+1}} > \left[l_1 L_2 + \Psi_2(\sigma)\right]$, i.e.,

$$\|\varsigma\|_2 > \left(\frac{l_1 L_2 + \Psi_2(\sigma)}{\lambda_{\min}\{Q\}-c_1}\right)^{\frac{\alpha+1}{2\alpha}} \quad (94)$$

the differential inequality (93) is finite time convergent. Because α is sufficiently small, $\frac{\alpha+1}{2\alpha}$ is sufficiently small, we want $\left(\frac{l_1 L_2 + \Psi_2(\sigma)}{\lambda_{\min}\{Q\}-c_1}\right)^{\frac{\alpha+1}{2\alpha}}$ to be sufficiently small, therefore, it is required that $0 < \frac{l_1 L_2 + \Psi_2(\sigma)}{\lambda_{\min}\{Q\}-c_1} < 1$.

Then, we have

$$c_1 < \lambda_{\min}\{Q\} - l_1 L_2 - \Psi_2(\sigma) \quad (95)$$

Therefore, from (88), we know that if

$$\|\varsigma\|_2 > \frac{\Psi_1(\sigma)}{\lambda_{\min}\{Q\} - l_1 L_2 - \Psi_2(\sigma)} \quad (96)$$

the differential inequality (93) is finite time convergent, and the error system (75) is finite-time stable. Therefore, we can get

$$\|\varsigma\|_2 \leq \frac{\Psi_1(\sigma)}{\lambda_{\min}\{Q\} - l_1 L_2 - \Psi_2(\sigma)} \quad (97)$$

This concludes the proof. ∎

## 6. Frequency analysis of continuous differentiator

For continuous differentiator (48), let $x_1 - v(t) = A\sin\omega t$, we have

$$\frac{2}{\pi}\int_0^\pi |A\sin\omega\tau|^{\frac{\alpha+1}{2}} \mathrm{sgn}(A\sin\omega\tau)\sin\omega\tau d\omega\tau \quad (98)$$
$$= A^{\frac{\alpha+1}{2}} \frac{2}{\pi}\int_0^\pi |\sin\omega\tau|^{\frac{\alpha+3}{2}} d\omega\tau$$

Denote $\Omega_1 = \frac{2}{\pi}\int_0^\pi |\sin\omega\tau|^{\frac{\alpha+3}{2}} d\omega\tau$, we have

$$1 < \Omega_1 < \frac{4}{\pi} \quad (99)$$

Therefore, the describing functions of nonlinear functions $|\ |^{\frac{\alpha+1}{2}}\mathrm{sgn}(\ )$ and $|\ |^\alpha \mathrm{sgn}(\ )$ are respectively

$$N_1(A) = A^{\frac{\alpha-1}{2}} \Omega_1 \quad (100)$$

and

$$N_2(A) = A^{\alpha-1}\Omega_2 \quad (101)$$

where

$$\Omega_2 = \frac{2}{\pi}\int_0^\pi |\sin\omega\tau|^{\alpha+1} d\omega\tau \quad (102)$$

Therefore, the linearization of continuous differentiator (48) is

$$\dot{x}_1 = x_2 - k_1 A^{\frac{\alpha-1}{2}} \Omega_1 (x_1 - v(t)) \quad (103)$$
$$\dot{x}_2 = -k_2 A^{\alpha-1} \Omega_2 (x_1 - v(t))$$

For system (103), the nature frequency is

$$\omega_n = \sqrt{k_2 \Omega_2} \Big/ A^{\frac{1-\alpha}{2}} \quad (104)$$

and from (103), we have

$$2\varsigma\omega_n = k_1 A^{\frac{\alpha-1}{2}} \Omega_1 \quad (105)$$

Therefore, the damping coefficient of (103) is

$$\varsigma = \frac{k_1 \Omega_1}{2\sqrt{k_2 \Omega_2}} \quad (106)$$

From (104), when the magnitude $A$ of tracking error is relatively small, $A^{\frac{1-\alpha}{2}} > A^{\frac{1}{2}} > A$, therefore, comparing with sliding mode differentiator, the nature frequency $\omega_n$ can be kept small. Moreover, the proposed differentiator is continuous, the outputs are all smooth. Therefore, the chattering phenomenon and high-frequency noises can be restrained sufficiently.

## 7. Simulations

In the following simulations, we select the function of $2\sin(t)$ as the signal $v(t)$.

### 7.1 Derivative estimation without noise
#### 7.1.1 Levant differentiator [7, 8]

$$\dot{x}_1 = x_2 - \lambda_2 |x_1 - v(t)|^{\frac{1}{2}} \mathrm{sgn}(x_1 - v(t))$$
$$\dot{x}_2 = -\lambda_1 \mathrm{sgn}(x_1 - v(t))$$

where $\lambda_2 = 6, \lambda_1 = 30$. Figures 3 and 4 show signal tracking and derivative estimation respectively by Levant differentiator without noise.

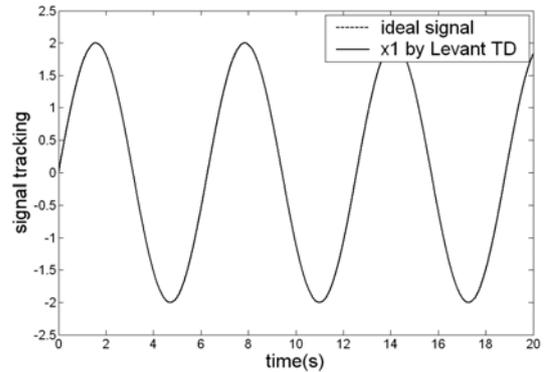

Fig. 3 Signal tracking by Levant differentiator

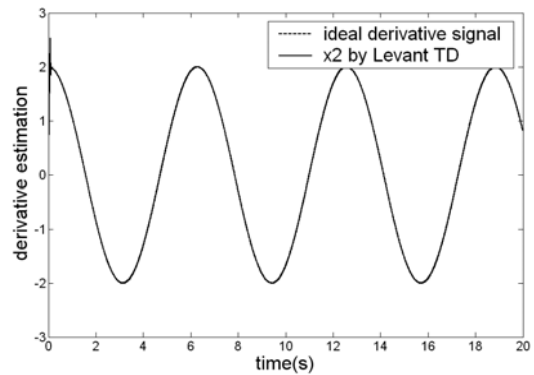

4-a. Derivative estimation by Levant differentiator

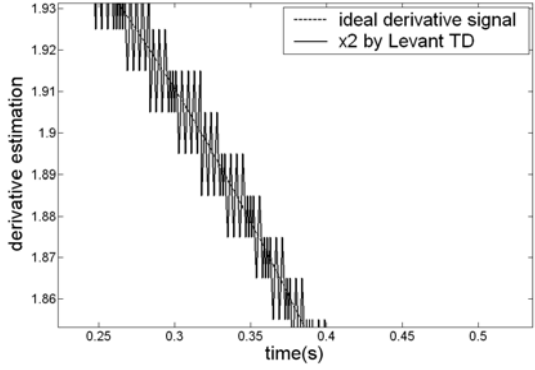

4-b The magnified figure of (a)

Fig. 4 Derivative estimation by Levant differentiator

Though the signal tracking output $x_1$ is smooth (See Fig. 3), derivative estimation output $x_2$ is continuous but non-smooth (See Fig. 4-b). The intensive chattering phenomenon happen due to discontinuous differentiator structure.

### 7.1.2 Continuous differentiator

$$\dot{x}_1 = x_2 - k_1 |x_1 - v(t)|^{\frac{\alpha+1}{2}} \operatorname{sgn}(x_1 - v(t))$$
$$\dot{x}_2 = -k_2 |x_1 - v(t)|^{\alpha} \operatorname{sgn}(x_1 - v(t))$$

where $k_1 = 6, k_2 = 30, \alpha = 0.2$. Figures 5 and 6 show signal tracking and derivative estimation respectively by continuous finite-time-convergent differentiator without noise.

From Fig. 5 and Fig. 6, signal tracking output $x_1$ and derivative estimation output $x_2$ are all smooth. The chattering phenomenon is avoided due to continuous differentiator structure.

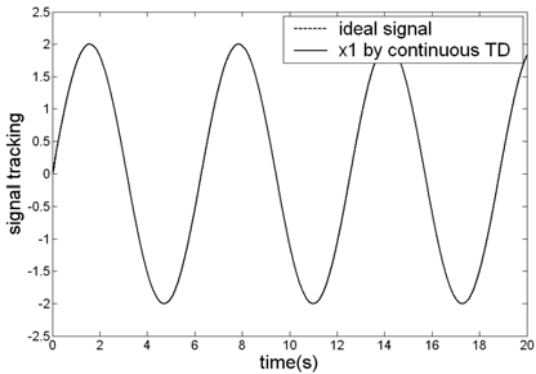

Fig. 5 Signal tracking by continuous differentiator

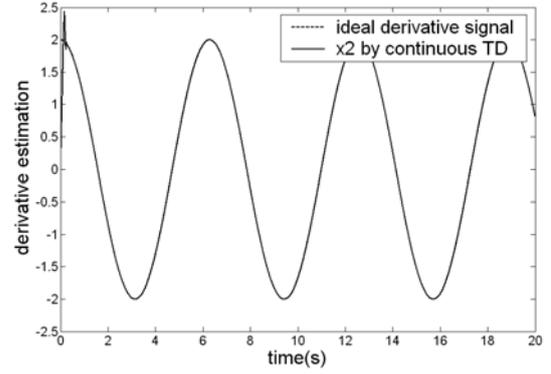

6-a Derivative estimation by continuous differentiator

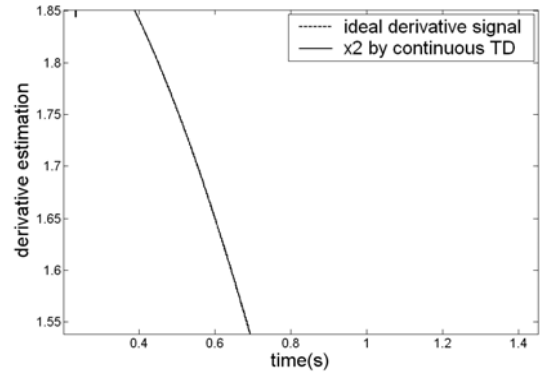

6-b The magnified figure of (a)

Fig. 6 Derivative estimation by continuous differentiator

## 7.2 Derivative estimation with noise
### 7.2.1 Levant differentiator [7, 8]

Parameters: $k_1 = 2, k_2 = 25$. Figures 7 and 8 show signal tracking and derivative estimation respectively by Levant differentiator with noises.

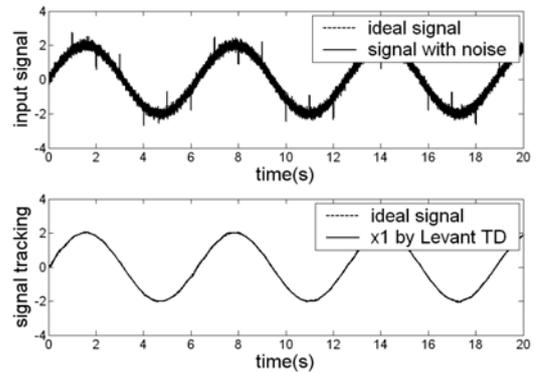

Fig. 7 Signal filtering and tracking by Levant differentiator

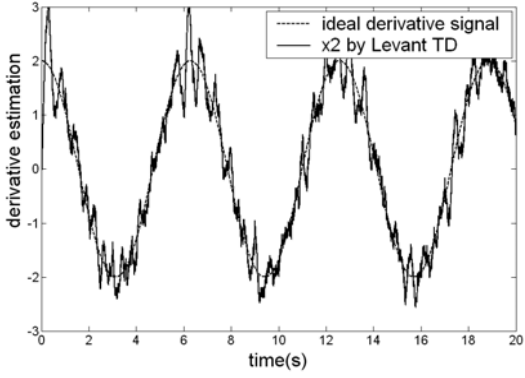

Fig. 8 Derivative estimation by Levant differentiator

From Fig. 8, we can find that there exists much noise in the derivative estimation output. Noise is magnified by chattering phenomenon near the equilibrium.

### 7.2.2 Continuous differentiator (48)

Parameters: $k_1 = 2, k_2 = 25, \alpha = 0.6$. Figures 9 and 10 show signal tracking and derivative estimation respectively by continuous finite-time-convergent differentiator (48) with noises.

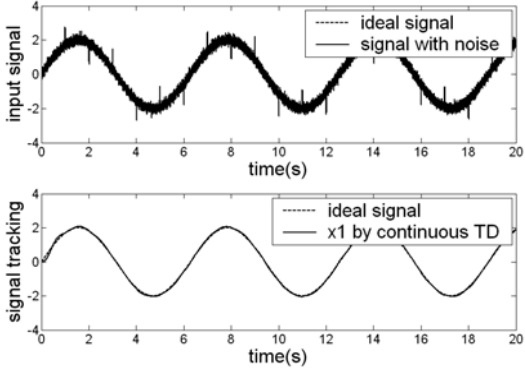

Fig. 9 Signal filtering and tracking by continuous differentiator

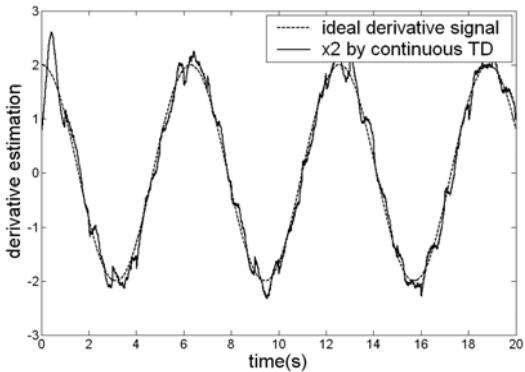

Fig. 10 Derivative estimation by continuous differentiator

From Fig. 9 and Fig. 10, we can find that signal tracking and derivative estimation are all satisfying in spite of the existence of much noise, and no time delay happen.

### 7.2.3 Finite-time convergent differentiator in [11]

$$\dot{x}_1 = x_2$$

$$\varepsilon^2 x_2 = -\left(x_1 - v(t) + \frac{3}{5}(\varepsilon x_2)^{\frac{5}{3}}\right)^{\frac{1}{5}} - (\varepsilon x_2)^{\frac{1}{3}}$$

where $\varepsilon = 0.1$. Figures 11 and 12 show signal tracking and derivative estimation respectively by differentiator in [11] with noises.

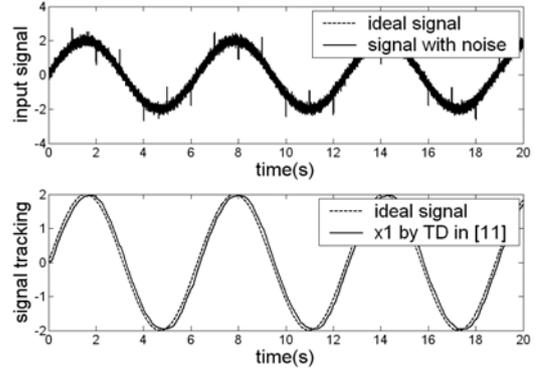

Fig. 11 Signal filtering and tracking by differentiator in [11]

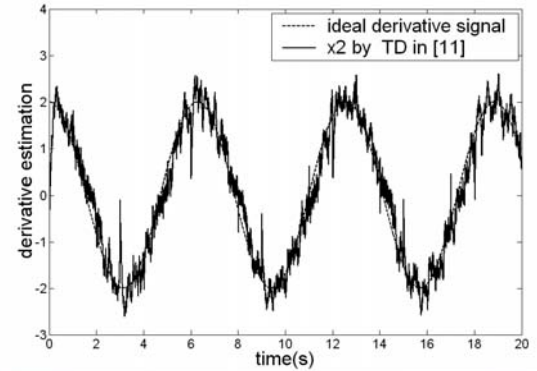

Fig. 12 Derivative estimation by differentiator in [11]

Without noise, the differentiator in [11] has a perfect tracking and estimation results (See the simulation figures in [11]). However, the filtering ability is far worse than that of continuous differentiator (48). Moreover, in order to restrain noise, the perturbation parameter ε should be selected larger than normal. Delay phenomenon exists in signal tracking because of the integral-chain structure in differentiator [11] (See Fig. 11), that is to say, there no lead compensation in integral-chain differentiator [11], but it exists in continuous differentiator (48).

From the simulations above, the presented continuous finite-time convergent differentiator has a better ability of restraining noises than sliding mode differentiator and the differentiator in [11].

## 8. Conclusion

In this paper, we present a continuous finite-time convergent differentiator based on a strong Lyapunov function. Because of its continuous structure, chattering phenomenon can be reduced sufficiently than sliding mode differentiator. Frequency analysis is applied to compare the continuous differentiator with sliding mode differentiator. Better restraining noise ability is obtained by this differentiator. Our future work is to design a robust differentiator for delayed signal and use harmonic analysis to optimize differentiators.